\documentclass[aps,prd,onecolumn,groupedaddress,showpacs,nofootinbib,amssymb]{revtex4}
\usepackage{graphicx,bm,color}
\usepackage{amsmath}
\usepackage{amssymb}
\usepackage{amsfonts}

\newcommand{\be}{\begin{equation}}
\newcommand{\ee}{\end{equation}}
\newcommand{\bea}{\begin{eqnarray}}
\newcommand{\eea}{\end{eqnarray}}
\newcommand{\beaa}{\begin{eqnarray*}}
\newcommand{\eeaa}{\end{eqnarray*}}

\newcommand{\nn}{\nonumber \\}
\newcommand{\e}{\mathrm{e}}
\newcommand{\tr}{\mathrm{tr}\,}

\allowdisplaybreaks[4]

\begin{document}

\title{Variety of cosmic acceleration models from  massive
$F(R)$ bigravity.}

\author{Shin'ichi Nojiri$^{1,2}$,
Sergei~D. Odintsov$^{3,4,5,6}$, and
Norihito Shirai$^1$
}

\affiliation{
$^1$ Department of Physics, Nagoya University, Nagoya
464-8602, Japan \\
$^2$ Kobayashi-Maskawa Institute for the Origin of Particles and
the Universe, Nagoya University, Nagoya 464-8602, Japan \\
$^3$ Consejo Superior de Investigaciones Cient\'{\i}ficas, ICE/CSIC-IEEC,
Campus UAB, Facultat de Ci\`{e}ncies, Torre C5-Parell-2a pl, E-08193
Bellaterra (Barcelona) Spain \\
$^4$ Instituci\'{o} Catalana de Recerca i Estudis Avan\c{c}ats (ICREA),
Barcelona, Spain\\
$^5$ Dept.Gen.\& Theor.Phys., Eurasian National Univ., Astana, Kazakhstan \\
$^6$ Tomsk State Pedagogical Univ., Tomsk, Russia
}

\begin{abstract}

We study the accelerating cosmology in massive $F(R)$ bigravity via the
reconstruction scheme. The consistent solution of the FRW equations is
presented:
it includes Big and Little Rip, quintessence, de Sitter and decelerating
universes described by the physical $g$ metric while the corresponding solution
of the universe described by the reference $f$ metric is also found.
It is demonstrated that in general the cosmological singularities of $g$ metric
are not always manifested as cosmological singularities of the reference $f$
metric.
We give two consistent ways to describe the Big and Little Rip, quintessence,
de Sitter and decelerating universes.
In one of the consistent solutions, the two metrics $g$ and $f$ coincide with
each other, which may indicate the connection with the convenient single 
metric background formulation.
For another solution, where  two metrics $g$ and $f$ do not coincide 
with each other, there always appears super-luminal mode.

\end{abstract}


\maketitle

\section{Introduction \label{SI}}

Massive gravity history has started from the work \cite{Fierz:1939ix}
(for recent review, see \cite{Hinterbichler:2011tt}).
It also has been known that massive gravity eventually contains
the Boulware-Deser ghost \cite{Boulware:1974sr}
and vDVZ discontinuity \cite{vanDam:1970vg} in the limit of $m\to 0$.
The attempts to screen this discontinuity are based on
the Vainstein mechanism \cite{Vainshtein:1972sx} as it was shown, for
instance, in Ref.~\cite{Luty:2003vm} on the example of the
DGP model \cite{Dvali:2000hr}.

It has been realized recently that non-linear massive gravity
\cite{deRham:2010ik,Hassan:2011hr} (with non-dynamical background metric)
may be extended to the ghost-free construction
with the dynamical metric \cite{Hassan:2011zd}.
The most general proof of absence of ghost in massive gravity has been given in
\cite{Hassan:2011tf}. 
Especially in case of the minimal model, which we treat in this paper in 
(\ref{bimetric2}) was first given in \cite{Hassan:2011vm}. 

General structure as well as cosmological evolution of such massive gravity has
been studied in Refs.~\cite{Golovnev:2011aa}.
The convenient description of the theory gives bigravity or bimetric gravity
which contains two metrics (symmetric tensor fields).
Some cosmological solutions of massive bigravity which describe accelerating
universe are known
\cite{Damour:2002wu,Volkov:2011an,vonStrauss:2011mq,Volkov:2012cf,Berg:2012kn,Nojiri:2012zu}.
One of two metrics is called physical metric while second metric is called
reference metric.
Despite the convenience, the interpretation of such description is not easy:
indeed, it looks like we have two different universes described by two
different metrics.
In reality, the second metric just gives the effective description of exotic
matter (massive graviton).

In Ref.~\cite{Nojiri:2012zu}, we have proposed  massive ghost-free
$F(R)$ bigravity which is formulated in terms of the auxiliary scalars.
The cosmological reconstruction scheme of $F(R)$ bigravity was
developed in detail with explicit examples of accelerating universe.
However, there may appear extra ghost (not of the Boulware-Deser type)
in such formulation.
Furthermore, as we show in this work, the scheme of the reconstruction
in Ref.~\cite{Nojiri:2012zu} is not fully consistent with the
Bianchi identities.

The present work is devoted to the study of accelerating cosmology in
massive $F(R)$ bigravity and its relation with observable universe.
%%%%%%%%%%%%%%
%The FRW cosmology in the standard bigravity models has been studied
%in \cite{} and references therein.
%%%%%%%%%%%%%%
After brief introduction to massive $F(R)$ bigravity in second section,
we develop the consistent reconstruction scheme where extra ghost is not
generated in the section III.
The qualitative difference with the case of usual F(R) gravity is caused 
by constraints which are provided by scalar equations and do not coincide 
with Bianchi identities.

We also study the relation between background cosmologies induced by two metrics.
We introduce four kinds of metrics, $g_{\mu\nu}$, $g^\mathrm{J}_{\mu\nu}$,
$f_{\mu\nu}$, and $f^\mathrm{J}_{\mu\nu}$.
The physical observable metric $g^\mathrm{J}_{\mu\nu}$ is the metric in the
Jordan frame.
The metric $g_{\mu\nu}$ corresponds to the metric in the Einstein frame in
the standard $F(R)$ gravity and therefore the metric $g_{\mu\nu}$ is not
physical metric.
In the bigravity theories, we have to introduce another reference metrics or
symmetric tensor $f_{\mu\nu}$ and $f^\mathrm{J}_{\mu\nu}$.
The metric $f_{\mu\nu}$ is the metric corresponding to the Einstein frame
with respect to the curvature given by the metric $f_{\mu\nu}$.
On the other hand, the metric  $f^\mathrm{J}_{\mu\nu}$ is the
metric corresponding to the Jordan frame.

Section IV is devoted to the search of accelerating cosmologies.
Variety of accelerating cosmologies which include Big Rip,
de Sitter, quintessence, and Little Rip universes are constructed
for the space-time described by the metric $g^\mathrm{J}_{\mu\nu}$
in the situation when the Einstein frame metric $g_{\mu\nu}$ is fixed.
These cosmologies are in good correspondence with observational data
which can be shown in the analogy with Ref.~\cite{Akrami:2012vf}.
It is demonstrated that in general, cosmological singularity in physical
universe given by the metric $g^\mathrm{J}_{\mu\nu}$ is manifested
in the universe given by the reference metric $f_{\mu\nu}$ or
$f^\mathrm{J}_{\mu\nu}$ and vice-versa (the corresponding observation
for $R$ massive gravity with matter is
presented in Ref.~\cite{Capozziello:2012re}).
However, we show that there are models where cosmological singularity does
not occur in the universe described by the metric $g^\mathrm{J}_{\mu\nu}$
although it occurs in the universe described by
the metric $f_{\mu\nu}$ or $f^\mathrm{J}_{\mu\nu}$.
It is presented the example where the universe given by the metric
$g^\mathrm{J}_{\mu\nu}$ accelerates while the universe given by the metric
$f_{\mu\nu}$ or $f^\mathrm{J}_{\mu\nu}$ decelerates.
In this sense, the space-time given by the metric $f_{\mu\nu}$ or
$f^\mathrm{J}_{\mu\nu}$ (i.e., massive graviton effect) plays a role of
dark energy in our universe, where the metric is given by
$g^\mathrm{J}_{\mu\nu}$.
We also propose the qualitative scenario of dissolution of the
background metric $f^\mathrm{J}_{\mu\nu}$ by the physical metric.
The explicit examples of Big Rip, quintessence, de Sitter, Little Rip or
decelerating universes as solutions of FRW equations in $F(R)$ bigravity
when $f$ metric coincides with $g$ metric are found by
using reconstruction scheme.
General arguments for future singularity occurrence
in the universe given by the metric $f^\mathrm{J}_{\mu\nu}$
but its avoidance in the physical universe given by $g^\mathrm{J}_{\mu\nu}$
are presented.
Section V is devoted to formulation of perturbation theory. It turns out 
that for the study of stability of cosmological solutions under discussion 
one should investigate the eigenvalues of eight by eight matrix.

Recently, the super-luminal mode in the massive gravity has been discussed
(see \cite{BezerradeMello:2012nq,%Deser:2012qx,
deRham:2011pt} and references 
therein).
In section VI, we also observed that the massless particle in the 
space-time given by the
metric $f_{\mu\nu}$ or $f^\mathrm{J}_{\mu\nu}$ can be super-luminal. Some 
physical consequences of this effect are briefly mentioned.
Finally, some summary and outlook are given in Discussion section.

\section{$F(R)$ bigravity \label{SII}}

In this section, we review the construction of ghost-free $F(R)$ bigravity,
following Ref.~\cite{Nojiri:2012zu} (for recent review of convenient
$F(R)$ gravity, see \cite{Nojiri:2006ri,Capozziello:2010zz}).
The consistent model of bimetric gravity, which includes two metric tensors
$g_{\mu\nu}$ and $f_{\mu\nu}$, was proposed in Ref.~\cite{Hassan:2011zd}.
It contains the massless spin-two field, corresponding to graviton, and
massive spin-two field.
The gravity model which only contains the massive spin-two field is called
massive gravity but we consider the model including both of massless and
massive spin two field, which is called bigravity.
It has been shown that the Boulware-Deser ghost \cite{Boulware:1974sr} does not
appear in such a theory.

The starting action is given by
\begin{align}
\label{bimetric}
S_\mathrm{bi} =&M_g^2\int d^4x\sqrt{-\det g}\,R^{(g)}+M_f^2\int d^4x
\sqrt{-\det f}\,R^{(f)} \nonumber \\
&+2m^2 M_\mathrm{eff}^2 \int d^4x\sqrt{-\det g}\sum_{n=0}^{4} \beta_n\,
e_n \left(\sqrt{g^{-1} f} \right) \, .
\end{align}
Here $R^{(g)}$ is the scalar curvature for $g_{\mu \nu}$ and
$R^{(f)}$ is the scalar curvature for $f_{\mu \nu}$.
$M_\mathrm{eff}$ is defined by
\be
\label{Meff}
\frac{1}{M_\mathrm{eff}^2} = \frac{1}{M_g^2} + \frac{1}{M_f^2}\, .
\ee
Furthermore, tensor $\sqrt{g^{-1} f}$ is defined by the square root of
$g^{\mu\rho} f_{\rho\nu}$, that is,
$\left(\sqrt{g^{-1} f}\right)^\mu_{\ \rho} \left(\sqrt{g^{-1}
f}\right)^\rho_{\ \nu} = g^{\mu\rho} f_{\rho\nu}$.
For general tensor $X^\mu_{\ \nu}$, $e_n(X)$'s are defined by
\begin{align}
\label{ek}
& e_0(X)= 1  \, , \quad
e_1(X)= [X]  \, , \quad
e_2(X)= \tfrac{1}{2}([X]^2-[X^2])\, ,\nn
& e_3(X)= \tfrac{1}{6}([X]^3-3[X][X^2]+2[X^3])
\, ,\nn
& e_4(X) =\tfrac{1}{24}([X]^4-6[X]^2[X^2]+3[X^2]^2
+8[X][X^3]-6[X^4])\, ,\nn
& e_k(X) = 0 ~~\mbox{for}~ k>4 \, .
\end{align}
Here $[X]$ expresses the trace of arbitrary tensor
$X^\mu_{\ \nu}$: $[X]=X^\mu_{\ \mu}$.

In order to construct the consistent $F(R)$ bigravity,
we add the following terms to the action (\ref{bimetric}):
\begin{align}
\label{Fbi1}
S_\varphi =& - M_g^2 \int d^4 x \sqrt{-\det g}
\left\{ \frac{3}{2} g^{\mu\nu} \partial_\mu \varphi \partial_\nu \varphi
+ V(\varphi) \right\} + \int d^4 x \mathcal{L}_\mathrm{matter}
\left( \e^{\varphi} g_{\mu\nu}, \Phi_i \right)\, ,\\
\label{Fbi7b}
S_\xi =& - M_f^2 \int d^4 x \sqrt{-\det f}
\left\{ \frac{3}{2} f^{\mu\nu} \partial_\mu \xi \partial_\nu \xi
+ U(\xi) \right\} \, .
\end{align}
By the conformal transformations
$g_{\mu\nu} \to \e^{-\varphi} g^{\mathrm{J}}_{\mu\nu}$ and
$f_{\mu\nu}\to \e^{-\xi} f^{\mathrm{J}}_{\mu\nu}$,
the total action $S_{F} = S_\mathrm{bi} + S_\varphi + S_\xi$
is transformed as
\begin{align}
\label{FF1}
S_{F} =& M_f^2\int d^4x\sqrt{-\det f^{\mathrm{J}}}\,
\left\{ \e^{-\xi} R^{\mathrm{J}(f)} -  \e^{-2\xi} U(\xi) \right\} \nn
& +2m^2 M_\mathrm{eff}^2 \int d^4x\sqrt{-\det g^{\mathrm{J}}}\sum_{n=0}^{4}
\beta_n
\e^{\left(\frac{n}{2} -2 \right)\varphi - \frac{n}{2}\xi} e_n
\left(\sqrt{{g^{\mathrm{J}}}^{-1} f^{\mathrm{J}}} \right) \nn
& + M_g^2 \int d^4 x \sqrt{-\det g^{\mathrm{J}}}
\left\{ \e^{-\varphi} R^{\mathrm{J}(g)} - \e^{-2\varphi} V(\varphi) \right\}
+ \int d^4 x \mathcal{L}_\mathrm{matter}
\left( g^{\mathrm{J}}_{\mu\nu}, \Phi_i \right)\, .
\end{align}
The kinetic terms for $\varphi$ and $\xi$ vanish. By the variations
with respect to $\varphi$ and $\xi$ as in the case of convenient $F(R)$
gravity \cite{Nojiri:2003ft}, we obtain
\begin{align}
\label{FF2}
0 =& 2m^2 M_\mathrm{eff}^2 \sum_{n=0}^{4} \beta_n \left(\frac{n}{2} -2 \right)
\e^{\left(\frac{n}{2} -2 \right)\varphi - \frac{n}{2}\xi} e_n
\left(\sqrt{{g^{\mathrm{J}}}^{-1} f^{\mathrm{J}}}\right)
+ M_g^2 \left\{ - \e^{-\varphi} R^{\mathrm{J}(g)} + 2  \e^{-2\varphi}
V(\varphi)
+ \e^{-2\varphi} V'(\varphi) \right\}\, ,\\
\label{FF3}
0 =& - 2m^2 M_\mathrm{eff}^2 \sum_{n=0}^{4} \frac{\beta_n n}{2}
\e^{\left(\frac{n}{2} -2 \right)\varphi - \frac{n}{2}\xi} e_n
\left(\sqrt{{g^{\mathrm{J}}}^{-1} f^{\mathrm{J}}}\right)
+ M_f^2 \left\{ - \e^{-\xi} R^{\mathrm{J}(f)} + 2  \e^{-2\xi} U(\xi)
+ \e^{-2\xi} U'(\xi) \right\}\, .
\end{align}
The  Eqs.~(\ref{FF2}) and (\ref{FF3}) can be solved algebraically
with respect to $\varphi$ and $\xi$ as
$\varphi = \varphi \left( R^{\mathrm{J}(g)}, R^{\mathrm{J}(f)},
e_n \left(\sqrt{{g^{\mathrm{J}}}^{-1} f^{\mathrm{J}}}\right)
\right)$ and
$\xi = \xi \left( R^{\mathrm{J}(g)}, R^{\mathrm{J}(f)},
e_n \left(\sqrt{{g^{\mathrm{J}}}^{-1}
f^{\mathrm{J}}}\right) \right)$.
Substituting above $\varphi$ and $\xi$ into (\ref{FF1}),
one gets  $F(R)$ bigravity:
\begin{align}
\label{FF4}
S_{F} =& M_f^2\int d^4x\sqrt{-\det f^{\mathrm{J}}}
F^{(f)}\left( R^{\mathrm{J}(g)}, R^{\mathrm{J}(f)},
e_n \left(\sqrt{{g^{\mathrm{J}}}^{-1} f^{\mathrm{J}}}\right) \right) \nn
& +2m^2 M_\mathrm{eff}^2 \int d^4x\sqrt{-\det g}\sum_{n=0}^{4} \beta_n
\e^{\left(\frac{n}{2} -2 \right)
\varphi\left( R^{\mathrm{J}(g)},
e_n \left(\sqrt{{g^{\mathrm{J}}}^{-1} f^{\mathrm{J}}}\right) \right)}
e_n \left(\sqrt{{g^{\mathrm{J}}}^{-1} f^{\mathrm{J}}} \right) \nn
& + M_g^2 \int d^4 x \sqrt{-\det g^{\mathrm{J}}}
F^{\mathrm{J}(g)}\left( R^{\mathrm{J}(g)}, R^{\mathrm{J}(f)},
e_n \left(\sqrt{{g^{\mathrm{J}}}^{-1} f^{\mathrm{J}}}\right) \right)
+ \int d^4 x \mathcal{L}_\mathrm{matter}
\left( g^{\mathrm{J}}_{\mu\nu}, \Phi_i \right)\, ,
\end{align}
\begin{align}
\label{FF4BBB}
F^{\mathrm{J}(g)}\left( R^{\mathrm{J}(g)}, R^{\mathrm{J}(f)},
e_n \left(\sqrt{{g^{\mathrm{J}}}^{-1} f^{\mathrm{J}}}\right) \right)
\equiv &
\left\{ \e^{-\varphi\left( R^{\mathrm{J}(g)}, R^{\mathrm{J}(f)},
e_n \left(\sqrt{{g^{\mathrm{J}}}^{-1} f^{\mathrm{J}}}\right)
\right)} R^{\mathrm{J}(g)} \right. \nn & \left.
 -  \e^{-2\varphi\left( R^{\mathrm{J}(g)}, R^{\mathrm{J}(f)},
e_n \left(\sqrt{{g^{\mathrm{J}}}^{-1} f^{\mathrm{J}}}\right)
\right)}
V \left(\varphi\left( R^{\mathrm{J}(g)}, R^{\mathrm{J}(f)},
e_n \left(\sqrt{{g^{\mathrm{J}}}^{-1} f^{\mathrm{J}}}\right)
\right)\right) \right\} \, ,\nn
F^{(f)}\left( R^{\mathrm{J}(g)}, R^{\mathrm{J}(f)},
e_n \left(\sqrt{{g^{\mathrm{J}}}^{-1} f^{\mathrm{J}}}\right) \right)
\equiv &
\left\{ \e^{-\xi\left( R^{\mathrm{J}(g)}, R^{\mathrm{J}(f)},
e_n \left(\sqrt{{g^{\mathrm{J}}}^{-1} f^{\mathrm{J}}}\right)
\right)} R^{\mathrm{J}(f)} \right. \nn
& \left.
 -  \e^{-2\xi\left( R^{\mathrm{J}(g)}, R^{\mathrm{J}(f)},
e_n \left(\sqrt{{g^{\mathrm{J}}}^{-1} f^{\mathrm{J}}}\right) \right)}
U \left(\xi\left( R^{\mathrm{J}(g)}, R^{\mathrm{J}(f)},
e_n \left(\sqrt{{g^{\mathrm{J}}}^{-1} f^{\mathrm{J}}}\right)
\right)\right) \right\} \, .
\end{align}
Note that it is difficult to solve Eqs.~(\ref{FF2}) and
(\ref{FF3}) with respect to $\varphi$ and $\xi$ explicitly.
Therefore, it might be easier to define the model
in terms of the auxiliary scalars $\varphi$ and $\xi$ as in (\ref{FF1}).

\section{Cosmological Reconstruction \label{SIII}}

Let us consider the cosmological reconstruction program
following Ref.~\cite{Nojiri:2012zu} but in slightly extended form.

For simplicity, we start from the minimal case
\begin{align}
\label{bimetric2}
S_\mathrm{bi} =&M_g^2\int d^4x\sqrt{-\det g}\,R^{(g)}+M_f^2\int d^4x
\sqrt{-\det f}\,R^{(f)} \nonumber \\
&+2m^2 M_\mathrm{eff}^2 \int d^4x\sqrt{-\det g} \left( 3 - \tr \sqrt{g^{-1} f}
+ \det \sqrt{g^{-1} f} \right)\, .
\end{align}
In order to evaluate $\delta \sqrt{g^{-1} f}$, two matrices $M$ and
$N$, which satisfy the relation $M^2=N$ are taken.
Since $\delta M M + M \delta M = \delta N$, one finds
\be
\label{Fbi7}
\tr \delta M = \frac{1}{2} \tr \left( M^{-1} \delta N \right)\, .
\ee
For a while, we consider the Einstein frame action (\ref{bimetric2}) with
(\ref{Fbi1}) and (\ref{Fbi7b}) but matter contribution is neglected.
Then by the variation over $g_{\mu\nu}$, we obtain
\begin{align}
\label{Fbi8}
0 =& M_g^2 \left( \frac{1}{2} g_{\mu\nu} R^{(g)} - R^{(g)}_{\mu\nu} \right)
+ m^2 M_\mathrm{eff}^2 \left\{ g_{\mu\nu} \left( 3 - \tr \sqrt{g^{-1} f}
\right)
+ \frac{1}{2} f_{\mu\rho} \left( \sqrt{ g^{-1} f } \right)^{-1\, \rho}_{\qquad 
\nu}
+ \frac{1}{2} f_{\nu\rho} \left( \sqrt{ g^{-1} f } \right)^{-1\, \rho}_{\qquad 
\mu}
\right\} \nn
& + M_g^2 \left[ \frac{1}{2} \left( \frac{3}{2} g^{\rho\sigma} \partial_\rho
\varphi \partial_\sigma \varphi
+ V (\varphi) \right) g_{\mu\nu} - \frac{3}{2}
\partial_\mu \varphi \partial_\nu \varphi \right] \, .
\end{align}
On the other hand, by the variation over $f_{\mu\nu}$, we get
\begin{align}
\label{Fbi9}
0 =& M_f^2 \left( \frac{1}{2} f_{\mu\nu} R^{(f)} - R^{(f)}_{\mu\nu} \right) \nn
& + m^2 M_\mathrm{eff}^2 \sqrt{ \det \left(f^{-1}g\right) } \left \{
  - \frac{1}{2}f_{\mu\rho} \left( \sqrt{g^{-1} f} \right)^{\rho}_{\ \nu} 
  - \frac{1}{2}f_{\nu\rho} \left( \sqrt{g^{-1} f} \right)^{\rho}_{\ \mu} 
+ \det \left( \sqrt{g^{-1} f} \right) f_{\mu\nu} \right\}
%\textcolor{red}{
%\det \sqrt{f^{-1}g} \left ( \sqrt{f^{-1}g} \right )^{-1 \rho}_{\qquad \mu} 
%f_{\rho \nu} - f_{\mu \nu}
%}
% \right \}
\nn
%m^2 M_\mathrm{eff}^2 \left\{ f_{\mu\nu}
%\left( 3 - \tr \sqrt{g^{-1} f} \right) - f_{\mu\sigma}
%\left( \sqrt{ g^{-1} f } \right)^{-1\, \sigma}_{\qquad \rho}
%g^{\rho\tau} f_{\tau\nu}\right\} \nn
& + M_f^2 \left[ \frac{1}{2} \left( \frac{3}{2} f^{\rho\sigma} \partial_\rho
\xi \partial_\sigma \xi
+ U (\xi) \right) f_{\mu\nu} - \frac{3}{2} \partial_\mu \xi \partial_\nu \xi
\right] \, .
\end{align}
We should note that $\det \sqrt{g} \det \sqrt{g^{-1} f } \neq \sqrt{f}$ in general.
%%%%%%%%%%
%%%%%%%%%%%
The variations of the scalar fields $\varphi$ and $\xi$ are given by
\be
\label{scalareq}
0 = - 3 \Box_g \varphi + V' (\varphi) \, ,\quad
0 = - 3 \Box_f \xi + U' (\xi) \, .
\ee
Here $\Box_g$ ($\Box_f$) is the d'Alembertian with respect to the metric $g$ 
($f$).
By multiplying the covariant derivative $\nabla_g^\mu$ with respect to the 
metric $g$ with
Eq.~(\ref{Fbi8}) and using the Bianchi identity
$0=\nabla_g^\mu\left( \frac{1}{2} g_{\mu\nu} R^{(g)} - R^{(g)}_{\mu\nu} 
\right)$ and
Eq.~(\ref{scalareq}), we obtain
\be
\label{identity1}
0 = - g_{\mu\nu} \nabla_g^\mu \left( \tr \sqrt{g^{-1} f} \right)
+ \frac{1}{2} \nabla_g^\mu \left\{ f_{\mu\rho} \left( \sqrt{ g^{-1} f } 
\right)^{-1\, \rho}_{\qquad \nu}
+ f_{\nu\rho} \left( \sqrt{ g^{-1} f } \right)^{-1\, \rho}_{\qquad \mu} 
\right\} \, .
\ee
Similarly by using the covariant derivative $\nabla_f^\mu$ with respect to the 
metric $f$, from (\ref{Fbi9}),
we obtain
\be
\label{identity2}
0 = \nabla_f^\mu \left[
\sqrt{ \det \left(f^{-1}g\right) } \left \{
  - \frac{1}{2}\left( \sqrt{g^{-1} f} \right)^{ -1 \nu}_{\ \ \ \ \ \sigma} 
g^{\sigma\mu}
  - \frac{1}{2}\left( \sqrt{g^{-1} f} \right)^{ -1 \mu}_{\ \ \ \ \sigma} 
g^{\sigma\nu}
+ \det \left( \sqrt{g^{-1} f} \right) f^{\mu\nu} \right\} \right]\, .
\ee
In case of the Einstein gravity, the conservation law of the energy-momentum
  tensor depends from the Einstein equation. It can be derived from the Bianchi 
identity.
In case of bigravity, however, the conservation laws of the energy-momentum 
tensor of the scalar fields
  are derived
from the scalar field equations. These conservation laws are independent of the 
Einstein equation. The
Bianchi identities give equations (\ref{identity1}) and (\ref{identity2}) 
independent of the Einstein equation.

We now assume the FRW universes for the metrics $g_{\mu\nu}$ and $f_{\mu\nu}$
and use the conformal time $t$ for the universe with metric $g_{\mu\nu}$
\footnote{
In Ref.~\cite{Nojiri:2012zu}, we have used the cosmological time instead of the
conformal time. The use of the conformal time simplifies the formulation.}:
\be
\label{Fbi10}
ds_g^2 = \sum_{\mu,\nu=0}^3 g_{\mu\nu} dx^\mu dx^\nu
= a(t)^2 \left( - dt^2 + \sum_{i=1}^3 \left( dx^i \right)^2\right) \, ,\quad
ds_f^2 = \sum_{\mu,\nu=0}^3 f_{\mu\nu} dx^\mu dx^\nu
= - c(t)^2 dt^2 + b(t)^2 \sum_{i=1}^3 \left( dx^i \right)^2 \, .
\ee
Then $(t,t)$ component of (\ref{Fbi8}) gives
\be
\label{Fbi11}
0 = - 3 M_g^2 H^2 - 3 m^2 M_\mathrm{eff}^2
\left( a^2 - ab \right) + \left(
%
%\textcolor{red}{+}
\frac{3}{4}
{\dot\varphi}^2
%-
%\textcolor{red}{+}
+ \frac{1}{2} V (\varphi) a(t)^2 \right) M_g^2 \, ,
\ee
and $(i,j)$ components give
\be
\label{Fbi12}
0 = M_g^2 \left( 2 \dot H + H^2 \right)
+  m^2 M_\mathrm{eff}^2 \left( 3a^2 - 2ab - ac \right) + \left(
% -
%\textcolor{red}{+}
\frac{3}{4} {\dot\varphi}^2
%+
%\textcolor{red}{-}
  -
\frac{1}{2} V (\varphi) a(t)^2 \right) M_g^2 \, .
\ee
Here $H=\dot a / a$.
On the other hand,  $(t,t)$ component of (\ref{Fbi9}) gives
\be
\label{Fbi13}
0 = - 3 M_f^2 K^2 +  m^2 M_\mathrm{eff}^2 c^2
%\left( - 3 + \frac{2c}{a} + \frac{3 b}{a} \right)
%\textcolor{red}{
\left ( 1 - \frac{a^3}{b^3} \right )
%}
+ \left(
%-
%\textcolor{red}{+}
\frac{3}{4} {\dot\xi}^2
%-
%\textcolor{red}{+}
  -
\frac{1}{2} U (\xi) c(t)^2 \right) M_f^2 \, ,
\ee
and $(i,j)$ components give
\be
\label{Fbi14}
0 = M_f^2
%
%\textcolor{red}{\frac{b^2}{c^2}}
%
\left( 2 \dot K + 3 K^2 - 2 LK \right)
+  m^2 M_\mathrm{eff}^2
%\textcolor{red}{
\left( \frac{a^3c}{b^2} - c^2
%3 - \frac{c}{a} - \frac{7b}{a}
\right)
%}
+ \left(
%-
%\textcolor{red}{+}
\frac{3}{4} {\dot\xi}^2
% +
%\textcolor{red}{-}
  -
\frac{1}{2} U (\xi) c(t)^2 \right) M_f^2 \, .
\ee
Here $K =\dot b / b$ and $L= \dot c / c$.
Both of Eq.~(\ref{identity1}) and Eq.~(\ref{identity2}) give the identical 
equation:
\be
\label{identity3}
cH = bK\ \mbox{or}\
\frac{c\dot a}{a} = \dot b\, .
\ee
If $\dot a \neq 0$, we obtain $c= a\dot b / \dot a$.
On the other hand, if $\dot a = 0$, we find $\dot b=0$, that is, $a$ and $b$ 
are
constant and $c$ can be arbitrary.
We should note that the scheme of the reconstruction in 
Ref.~\cite{Nojiri:2012zu},
where $a(t)=c(t)=1$ is not, unfortunately, consistent with 
Eq.~(\ref{identity3})
in general.

We now redefine scalars as $\varphi=\varphi(\eta)$ and
$\xi = \xi (\zeta)$ and
identify $\eta$ and $\zeta$ with the conformal time $t$, $\eta=\zeta=t$.
Hence, one gets
\begin{align}
\label{Fbi19}
\omega(t) M_g^2 =&
%4 M_g^2 \left( \dot H - H^2 \right) + m^2 M_\mathrm{eff}^2
%\left( - a c + ab\right)
%\textcolor{red}{
  -4M_g^2 \left ( \dot{H}-H^2  \right )-2m^2 M^2_\mathrm{eff}(ab-ac)
%}
  \, , \\
\label{Fbi20}
\tilde V (t) a(t)^2 M_g^2 =&
%\textcolor{red}{
M_g^2 \left (2 \dot{H}+4 H^2 \right ) +m^2 M^2_\mathrm{eff}(6a^2-5ab-ac)
%}
% - M_g^2 \left( 2 \dot H + 4 H^2 \right) - m^2
%M_\mathrm{eff}^2 \left( 6 a^2  - 5ab - ac \right)
\, , \\
\label{Fbi21}
\sigma(t) M_f^2 =&
%4 M_f^2 \left( \dot K - L K \right) + 2 m^2 M_\mathrm{eff}^2
%c^2 \left( \frac{c}{a}  - \frac{b}{a} \right)
%\textcolor{red}{
  - 4 M_f^2 \left ( \dot{K} - LK  \right )
  - 2m^2 M_\mathrm{eff}^2 \left ( - \frac{c}{b} + 1 \right ) \frac{a^3c}{b^2}
%}
\, , \\
\label{Fbi22}
\tilde U (t) c(t)^2 M_f^2 =&
%\textcolor{red}{
M_f^2 \left ( 2 \dot{K} + 6 K^2 -2 L K  \right )
+ m^2 M_\mathrm{eff}^2 \left( \frac{a^3c}{b^2} - 2 c^2 + \frac{a^3c^2}{b^3} 
\right)
% }
% - M_f^2 \left( 2 \dot K + 6 K^2 - 2 L K \right) -
%m^2 M_\mathrm{eff}^2 c^2 \left( 6 - \frac{3c}{a}  - \frac{7 b}{a} \right)
\, .
\end{align}
Here
\be
\label{Fbi23}
\omega(\eta) = 3 \varphi'(\eta)^2 \, ,\quad
\tilde V(\eta) = V\left( \varphi\left(\eta\right) \right)\, ,\quad
\sigma(\zeta) = 3 \xi'(\zeta)^2 \, ,\quad
\tilde U(\zeta) = U \left( \xi \left(\zeta\right) \right) \, .
\ee
Therefore for arbitrary $a(t)$, $b(t)$, and $c(t)$ if we choose $\omega(t)$,
$\tilde V(t)$, $\sigma(t)$, and $\tilde U(t)$
to satisfy Eqs.~(\ref{Fbi19}-\ref{Fbi22}), the cosmological model with given
$a(t)$, $b(t)$ and $c(t)$ evolution can be reconstructed.

\section{Accelerating cosmological models \label{SIV}}

Let us construct some examples of cosmological models
which describe Big Rip universe \cite{Caldwell:1999ew,Caldwell:2003vq},
quintessence universe, de Sitter universe, decelerating universe,
and Little Rip universe
\cite{Frampton:2011sp,Brevik:2011mm,Frampton:2011rh}.
The physical metric, where the scalar does not directly
couple with matter, is given by multiplying the scalar field to the metric in
the Einstein frame in (\ref{Fbi1}) or (\ref{bimetric2}):
\be
\label{Fbi30b}
g^\mathrm{J}_{\mu\nu} = \e^{\varphi} g_{\mu\nu}\, .
\ee
In the bigravity model, there appears another (unphysical) metric tensor
$f_{\mu\nu}$ besides $g_{\mu\nu}$.
In our model, since the matter only couples with $g_{\mu\nu}$, the physical
metric could be given by $g^\mathrm{J}_{\mu\nu}$ in (\ref{Fbi30b}).
In principle, however, there could be a matter
coupled with $f_{\mu\nu}$.
Then one may consider two space-times, one is described by $g_{\mu\nu}$ (or
$g^\mathrm{J}_{\mu\nu}$) and another by $f_{\mu\nu}$.
Although the matter in the space-time described by $f_{\mu\nu}$ could not
directly couple with the matter in the space-time described by $g_{\mu\nu}$,
the matters can interact with each other via the propagation of massless and
massive gravitons.
Furthermore, as clear from Eqs.~(\ref{Fbi8}) and (\ref{Fbi9}), the metric
$f_{\mu\nu}$ affects the geometry
of the space-time described by $g_{\mu\nu}$, and vice-versa.

In this section, we show that in general, there occurs future
singularity in the space-time
described by $g_{\mu\nu}$ when the singularity occurs in the space-time
described by $f_{\mu\nu}$.
We also find some examples where the future singularity does not appear in
universe described by $g_{\mu\nu}$ (more exactly by $g^\mathrm{J}_{\mu\nu}$)
even if future singularity appears in universe described by $f_{\mu\nu}$.
This might be a counterexample for the observation of
Ref.~\cite{Capozziello:2012re} where $R$ bigravity with matter was considered.
Although the physical metric $g^\mathrm{J}_{\mu\nu}$ in $g$ universe is
given by (\ref{Fbi30b}), it is not so clear what could be a physical metric in
$f$
universe since it depends on the coupling with matter.
Anyway we may consider both of $f_{\mu\nu}$ in the Einstein frame and the
metric
\be
\label{FFFFbi1}
f^\mathrm{J}_{\mu\nu} = \e^{\xi} f_{\mu\nu}\, ,
\ee
in the Jordan frame.

\subsection{Conformal description of accelerating universe \label{subVI1}}

In our formulation, it is convenient to use the conformal time description.
Hence, let us describe how the known cosmologies can be expressed by using the
conformal time.
Especially we present the explicit expressions of the de Sitter, phantom,
quintessence, decelerating, and also Little Rip universes.

The conformally flat FRW universe metric is given by
\be
\label{Fbi31}
ds^2 = \tilde a(t)^2 \left( - dt^2 + \sum_{i=1}^3 \left( dx^i \right)^2
\right) \, .
\ee
Eq.~(\ref{Fbi30b}) with (\ref{Fbi31}) shows
\be
\label{FFbi1}
\e^{\varphi(t)} a(t)^2 = \tilde a(t)^2 \, ,
\ee
that is,
\be
\label{FFbi2}
\varphi = - 2 \ln a(t) + %{\color{red}
2%}
\ln \tilde a(t)\, .
\ee
Using (\ref{Fbi23}), we find
\be
\label{FFbi3}
\omega(t) = 12 \left( H - \tilde H \right)^2 \, .
\ee
Here $\tilde H \equiv \frac{1}{\tilde a}\frac{d\tilde a}{dt}$.

In Eq.~(\ref{Fbi31}), when $\tilde a(t)^2 = \frac{l^2}{t^2}$, the metric
(\ref{Fbi31}) corresponds to the de Sitter universe which may describe
inflation or dark energy in the model under consideration.
On the other hand if $\tilde a(t)^2 = \frac{l^{2n}}{t^{2n}}$ with $n\neq 1$, by
redefining the time coordinate as
\be
\label{Fbi32}
d\tilde t = \pm \frac{l^n}{t^n}dt\, ,
\ee
that is,
\be
\label{Fbi32b}
\tilde t = \pm \frac{l^n}{n-1} t^{1-n}\, ,
\ee
the metric (\ref{Fbi31}) can be rewritten as
\be
\label{Fbi33}
ds^2 =  - d{\tilde t}^2 + \left( \pm (n-1) \frac{\tilde t}{l}
\right)^{- \frac{2n}{1-n}} \sum_{i=1}^3 \left( dx^i \right)^2  \, .
\ee
Eq.~(\ref{Fbi33}) shows that if $0<n<1$, the metric corresponds to the phantom
universe, if $n>1$ to the quintessence universe, and if $n<0$ to decelerating
universe.
In case of the phantom universe ($0<n<1$), one should choose $+$ sign in $\pm$
of (\ref{Fbi32}) or (\ref{Fbi32b}) and shift $\tilde t$ in (\ref{Fbi33})
as $\tilde t\to \tilde t - t_0$.
Then $\tilde t=t_0$ corresponds to the Big Rip and the present time is
$\tilde t<t_0$ and $t\to\infty$ corresponds to the infinite past
($\tilde t\to - \infty$).
In case of the quintessence universe ($n>1$), we may again choose $+$ sign in
$\pm$ of (\ref{Fbi32}) or (\ref{Fbi32b}).
Then $t\to 0$ corresponds to $\tilde t\to + \infty$ and $t\to +\infty$ to
$\tilde t \to 0$, which may correspond to the Big Bang.
In case of the decelerating universe ($n<0$), we may choose $-$ sign in $\pm$
of (\ref{Fbi32}) or (\ref{Fbi32b}).
Then $t\to 0$ corresponds to $\tilde t\to + \infty$ and $t\to +\infty$ to
$\tilde t \to 0$, which may correspond to the Big Bang, again.
We should also note that in case of the de Sitter universe ($n=1$), $t\to 0$
corresponds
to $\tilde t \to + \infty$ and $t\to \pm \infty$ to $\tilde t \to - \infty$.

One may also consider the Little Rip universe, where there is no future
singularity
but the Hubble rate in terms of the cosmological time $\tilde t$ becomes
infinite when $\tilde t$ goes to infinity.
As the universe expands, the relative acceleration between two points separated
by a comoving distance $l$ is given by
$l \left(1/\tilde a\right) \left(d^2 \tilde a/ d{\tilde t}^2\right)$,
where $a$ is the scale factor.
An observer at comoving distance $l$ away from a mass $m$ will measure an
inertial force on the mass of
\be
\label{i1}
F_\mathrm{iner}= \frac{m l}{\tilde a} \frac{d^2 \tilde a}{d{\tilde t}^2}\, .
\ee
Let us assume the
two particles are bound by a constant force $F_0$. If $F_\mathrm{iner}$ is
positive and greater
than $F_0$, the two particles become unbound. This is the ``rip'' produced by
the accelerating expansion.
It leads to finite-time disintegration of bound objects much before
the singularity.

An example is given by
\be
\label{LR1}
\frac{1}{\tilde a} \frac{d \tilde a}{d \tilde t}
= g_0 \lambda \e^{\lambda \tilde t} - \lambda \, .
\ee
The last term $-\lambda$ is added for the convenience in the explicit
calculation
but this term can be neglected for large $\tilde t$ compared with the first
term.
By the results of the Supernova Cosmology Project \cite{Amanullah:2010vv},
the parameter $\lambda$ in (\ref{LR1}) is bounded as
\be
\label{bound}
2.37\times 10^{-3}\, \mathrm{Gyr}^{-1}< \lambda < 8.37 \times 10^{-3}\,
\mathrm{Gyr}^{-1}\, .
\ee
Eq.~(\ref{LR1}) gives
\be
\label{LR2}
t = \frac{\e^{- g_0  \e^{\lambda \tilde t}}}{\lambda g_0}\, ,\quad
\tilde a = \e^{g_0 \e^{\lambda \tilde t} - \lambda \tilde t}
= - \frac{1}{\lambda t \ln \left( \lambda g_0 t\right) }\, .
\ee
By using (\ref{i1}), we find the inertial force is given by,
\be
\label{LR1b}
F_\mathrm{iner}=m l \left\{ g_0 \lambda^2 \e^{\lambda \tilde t}
+ \left( g_0 \lambda \e^{\lambda \tilde t}
- \lambda \right)^2 \right\}\, ,
\ee
which goes to infinity at infinite $\tilde t$.

Combining (\ref{LR1}) and (\ref{LR2}), one finds
\be
\label{LR3}
\tilde H = \frac{1}{\tilde a} \frac{d \tilde a}{d t}
= - \frac{1}{t} \left( 1 + \frac{1}{\ln \left( \lambda g_0 t \right)}
\right)\, .
\ee
Thus we get an explicit example of $\tilde a$ and $H$ corresponding to
the realistic Little Rip universe.

If the space-time described by the metric $g^\mathrm{J}_{\mu\nu}$ describes
the universe where we live, the functions $c(t)$ and $b(t)$ are
not directly related with the expansion of our universe since the functions
$c(t)$ and $b(t)$ correspond to the degrees of freedom in the Einstein frame
metric $f_{\mu\nu}$.

Therefore one may choose
$c(t)$ and $b(t)$ in the consistent way convenient for the calculation. This
does not mean
$c(t)$ and $b(t)$ are not relevant for the physics besides the expansion
of our universe since the matter in the universe given by the metric
$f_{\mu\nu}$ or $f^\mathrm{J}_{\mu\nu}$, if any, weakly interacts
with the matter in our universe via the massless and massive gravitons.
In the following, we consider two choices of $c(t)$ and $b(t)$, that is,
$a(t)=c(t)=1$ case and $a(t)=c(t)=b(t)$ case. Note that the last case probably
helps to simplify the formal description of the theory: indeed reference metric
seems to be dissolved by physical one in such situation.

\subsection{Dark energy universe with $a(t)=b(t)=1$ \label{subVI2_0}}

In this section, making the choice $a(t)=b(t)=1$, we explicitly construct
Big Rip (phantom), quintessence, de Sitter, decelerating or Little Rip
universes.
We should note that the choice $a(t)=b(t)=1$ satisfies the constraint 
(\ref{identity3}).

When $a(t)=b(t)=1$, the Einstein frame metric $g_{\mu\nu}$
expresses the flat Minkowski space although the metric we observe is given by
$g^\mathrm{J}_{\mu\nu}$.
Eqs.~(\ref{Fbi19}), (\ref{Fbi20}), (\ref{Fbi21}), and (\ref{Fbi22}) with
(\ref{FFbi3}) are simplified as follows,
\begin{align}
\label{Fbi19C}
\omega (t)^2 M_g^2 =& 12 M_g^2 \tilde H^2 =%& {\color{red}2%}
m^2 M_\mathrm{eff}^2
\left( c - 1\right) \, , \\
\label{Fbi20C}
\tilde V (t) M_g^2 =& m^2 M_\mathrm{eff}^2 \left( 1 - c \right)
= - 6 M_g^2 \tilde H^2 \, , \\
\label{Fbi21C}
\sigma(t) M_f^2 =& %{\color{red} -2 M_f^2 \left\{ 2b^2 \dot K + 3 \left(b^2 - 1 
%\right) K^2\right\}
2 m^2 M_\mathrm{eff}^2 \left( c - 1 \right)
%}
= 12 M_g^2 \tilde H^2 \, , \\
\label{Fbi22C}
\tilde U (t) M_f^2 =& %{\color{red}  M_f^2  \left\{ 2b^2 \dot K + 3 \left(b^2 + 
%1 \right) K^2\right\}
m^2 M_\mathrm{eff}^2 c \left( 1 - c \right) %}
= - 6 M_g^2 \tilde H^2
\left( 1 + \frac{ 6 \tilde H^2}{m^2 M_\mathrm{eff}^2} \right) \, .
\end{align}
Eq.~(\ref{Fbi19C}) can be solved with respect to $c$ as
\be
\label{FFFbi1}
c = 1 + \frac{ 6 \tilde H^2}{m^2 M_\mathrm{eff}^2} \, .
%b = 1 {\color{red} -} \frac{{\color{red} 6} M_g^2 }{m^2 
%M_\mathrm{eff}^2}\tilde H^2\, ,
\ee
We should note that both of $\omega(t)$ and $\sigma(t)$ are positive,
there does not appear ghost in the theory.

\subsubsection{Construction of the models describing Big Rip,
quintessence, de Sitter and decelerating universes \label{subsubVI1a}}

As shown in Subsection~\ref{subVI1}, Big Rip,
quintessence, de Sitter and  decelerating universes
are described by the scale factor
$\tilde a(t)^2 = \frac{l^{2n}}{t^{2n}}$ in terms of
the conformal time $t$.
Let us construct the models with the scale factor
$\tilde a(t)^2 = \frac{l^{2n}}{t^{2n}}$, that is $\tilde H = \frac{n}{t}$.
Studying the properties of such
models, we show that there does not appear future singularity in the space-time
described by $f^\mathrm{J}_{\mu\nu}$ although a future singularity appears
in the space-time described by $g^\mathrm{J}_{\mu\nu}$.

By using (\ref{Fbi19C}), (\ref{Fbi20C}), (\ref{Fbi21C}), and (\ref{Fbi22C}),
we find
\be
\label{FbiRRRA1}
\omega (t)^2 M_g^2 = \frac{12 n^2 M_g^2}{t^2} \, , \quad
\tilde V (t) M_g^2 = - \frac{6 n^2 M_g^2}{t^2} \, , \quad
\sigma(t) M_f^2 = \frac{12 n^2 M_g^2}{t^2} \, , \quad
\tilde U (t) M_f^2 =
= - \frac{6 n^2 M_g^2}{t^2} \left( 1 + \frac{ 6 n^2 }{m^2 M_\mathrm{eff}^2 t^2} 
\right) \, .
\ee
Eq.~(\ref{FbiRRRA1}) and $\sigma(t)$ in (\ref{Fbi23}) indicate
\be
\label{FFFF4}
\e^\xi = \frac{n^2}{t^2}\, .
\ee
Then by using (\ref{FFFbi1}),
`physical' metric $f^\mathrm{J}_{\mu\nu}$ in (\ref{FFFFbi1}) is given by
\be
\label{FFFF5}
\left( ds^\mathrm{J}_f \right)^2
= \sum_{\mu,\nu=0}^3 f^\mathrm{J}_{\mu\nu} dx^\mu dx^\nu
= \e^{\xi} ds_f^2
= \frac{n^2}{t^2} \left\{ - \left( 1 + \frac{ 6 n^2 }{m^2 M_\mathrm{eff}^2 
t^2}\right)^2
dt^2 + \left( dx^i \right)^2 \right\} \, .
\ee
When $t\sim 0$, by defining
\be
\label{FbiRRRA2}
\tilde t \sim \frac{\alpha}{2t^2}\, ,\quad
\alpha \equiv \frac{ 6 n^3 }{m^2 M_\mathrm{eff}^2 t^2}\, ,
\ee
we find  the metric (\ref{FFFF5})
\be
\label{FbiRRRA3}
\left( ds^\mathrm{J}_f \right)^2 \sim - d{\tilde t}^2
+ \frac{ 2 n^2 \tilde t}{\alpha} \left( dx^i \right)^2\, .
\ee
Because Eq.~(\ref{FbiRRRA2}) shows that $t\to 0$ corresponds to $\tilde t \to + 
\infty$,
there does not occur singularity in the metric $\left( ds^\mathrm{J}_f 
\right)^2$ because the scale factor
$\tilde a$ which is proportional to $\tilde t$ corresponds to the universe 
filled with radiation.

In summary, we presented the model where there does not occur cosmological 
singularity
in the universe described by $f^\mathrm{J}_{\mu\nu}$ but
there occurs finite-time future singularity in the universe described
by $g^\mathrm{J}_{\mu\nu}$.

\subsubsection{Little Rip universe \label{subsubVI1b}}

Let us discuss Little Rip universe which is realistic description of current 
universe. It may be consistent with observational bounds for LCDM as it was 
demonstrated earlier. In this section we show that the Little Rip cosmology in 
the space-time described by the
metric $g^\mathrm{J}_{\mu\nu}$ corresponds to the radiation dominated universe in the
space-time described by the metric $f^\mathrm{J}_{\mu\nu}$.
By substituting (\ref{LR2}) and (\ref{LR3}) into
(\ref{Fbi19C}), (\ref{Fbi20C}), (\ref{Fbi21C}), and (\ref{Fbi22C}),
we find
\begin{align}
\label{FbiRRRA4}
\omega (t)^2 M_g^2 =&
\frac{12 M_g^2}{t^2} \left( 1 + \frac{1}{\ln \left( \lambda g_0 t \right)}
\right)^2 \, , \quad
\tilde V (t) M_g^2 = - \frac{6 M_g^2}{t^2} \left( 1 + \frac{1}{\ln \left( 
\lambda g_0 t \right)}
\right)^2 \, , \nn
\sigma(t) M_f^2 =& \frac{12 M_g^2}{t^2} \left( 1 + \frac{1}{\ln \left( \lambda 
g_0 t \right)}
\right)^2 \, , \quad
\tilde U (t) M_f^2 = - \frac{6 M_g^2}{t^2} \left( 1 + \frac{1}{\ln \left( 
\lambda g_0 t \right)}
\right)^2\left\{ 1 +
\frac{6}{m^2 M_\mathrm{eff}^2 t^2} \left( 1 + \frac{1}{\ln \left( \lambda g_0 t 
\right)}
\right)^2 \right\}\, .
\end{align}
Then the metric $f^\mathrm{J}_{\mu\nu}$ in (\ref{FFFFbi1}) is given by
\be
\label{FbiRRRA5}
\left( ds^\mathrm{J}_f \right)^2
= \frac{1}{\left( \lambda t \ln \left( \lambda g_0 t\right) \right)^2}
\left[ - \left\{ 1 +
\frac{6}{m^2 M_\mathrm{eff}^2 t^2} \left( 1 + \frac{1}{\ln \left( \lambda g_0 t 
\right)}
\right)^2 \right\}^2 dt^2 + \left( dx^i \right)^2 \right] \, .
\ee
When $t$ is small, the metric behaves as,
\be
\label{FbiRRRA6}
\left( ds^\mathrm{J}_f \right)^2
\sim \frac{1}{\left( \lambda t \ln \left( \lambda g_0 t\right) \right)^2}
\left[ - \left(\frac{6}{m^2 M_\mathrm{eff}^2 t^2}\right)^2  dt^2 + \left( dx^i 
\right)^2 \right] \, .
\ee
We now define a new variable $\tilde t$ by
\be
\label{FbiRRRA7}
\tilde t \equiv \tilde \alpha \int \frac{dt}{t^3 \ln \left( \lambda g_0 
t\right) }
= - \frac{2\tilde \alpha}{t^2 \ln \left( \left( \lambda g_0 t\right)^2 \right)}
\sum_{n=0}^\infty \frac{n!}{ \left(- \ln \left( \left( \lambda g_0 t\right)^2 
\right)
\right)^n}\, ,
\quad
\tilde \alpha \equiv \frac{6}{\lambda m^2 M_\mathrm{eff}^2 } \, .
\ee
Then when $t\to 0+$, we have $\tilde t \to + \infty$ and
\be
\label{FbiRRRA8}
\left( \lambda g_0 t\right)^2 \sim \frac{2 \tilde \alpha}{\tilde t
\ln \left( \frac{\tilde t}{\left( \lambda g_0 \right)^2}\right)}\, .
\ee
Then the metric in (\ref{FbiRRRA6}) behaves as
\be
\label{FbiRRRA9}
\left( ds^\mathrm{J}_f \right)^2
\sim - d\tilde t^2 + \frac{g_0^2 \tilde t}{\alpha}\left( dx^i \right)^2 \, .
\ee
The asymptotic behavior of such the universe is identical with the universe 
filled by radiation.

\subsection{Dark energy universe with $a(t)=b(t)=c(t)$ \label{subVI2}}

As we observed above, general bigravity formally describes two gravities which
are related via some coupling term.
Generally speaking, this is rather formal presentation as second universe
described by reference metric is kind of effective description of exotic matter
(massive graviton).
Nevertheless, it may be useful to clarify the role of reference metric in the 
better way.
To do this, we propose that in the course of
the evolution the second universe metric may become equal to the physical
universe metric (of course, perturbation theories are different).
In other words, $f$ metric is dissolved by $g$ metric.
After that the future background evolution is conveniently described by the
single metric object.

Let us choose $a(t)=c(t)=b(t)$, which satisfy the condition (\ref{identity3}),
and therefore $H=K=L$. From (\ref{Fbi19}) and (\ref{Fbi21}),
we find $\omega(t)=\sigma(t)$ and
therefore $\varphi(t)=\xi(t)$, and also $V(t)= U(t)$ from (\ref{Fbi20}) and 
(\ref{Fbi22}),
which tells, not only $g_{\mu\nu}=f_{\mu\nu}$,
but $g^\mathrm{J}_{\mu\nu}=f^\mathrm{J}_{\mu\nu}$ from (\ref{Fbi30b}) and
(\ref{FFFFbi1}).
Hence if there is any singularity in the space-time described by $f_{\mu\nu}$
or $f^\mathrm{J}_{\mu\nu}$, there appears an identical singularity
in the universe described by $g_{\mu\nu}$ or $g^\mathrm{J}_{\mu\nu}$.
Note that, for the choice $a(t)=c(t)=b(t)$ in this Subsection,
there does  appear the ghost as it will be shown below.

By choosing $a(t)=c(t)=b(t)$,
Eqs.~(\ref{Fbi19}), (\ref{Fbi20}), (\ref{Fbi21}), and (\ref{Fbi22}) are
simplified as
\begin{align}
\label{Fbi19B}
3 \left( H - \tilde H \right)^2 =& - \dot H + H^2 \, , \\
\label{Fbi20B}
\tilde V (t) a(t)^2 =& \left( 2 \dot H + 4 H^2 \right) \, , \\
\label{Fbi21B}
\sigma(t)  =& 4 \left( - \dot H + H^2 \right) \, , \\
\label{Fbi22B}
\tilde U (t) a(t)^2 =& \left( 2 \dot H + 4 H^2 \right) \, .
\end{align}
Here, (\ref{FFbi3}) is used.
Comparing (\ref{Fbi19B}) with (\ref{Fbi21B}), we find
\be
\label{FFbi4}
\sigma(t) = \omega(t) = 4 \left( H - \tilde H \right)^2 \, .
\ee
Since $\sigma(t)$ is positive, there does not appear ghost.
In order to determine $a$, however, one should solve the differential
equation (\ref{Fbi19B}). In case if it has no solution, we cannot reconstruct
such a model.

\subsubsection{Construction of Big Rip, quintessence, de Sitter and
decelerating universes \label{subsubVI2a}}

Let us consider the construction of the models, which describe Big Rip,
quintessence, de Sitter and  decelerating universes, where the scale factor is
given by $\tilde a(t)^2 = \frac{l^{2n}}{t^{2n}}$.

In case $\tilde a(t)^2 = \frac{l^{2n}}{t^{2n}}$, the solution of
(\ref{Fbi19B}) is given by
\be
\label{FFbi6}
H = \frac{h_0}{t}\, ,\quad %{\color{red}
h_0 = - 2 + 3n \pm \sqrt{\left(2 - 3n\right)^2+ n^2}
%}
\, .
\ee
%{\color{red}
We should note that $h_0$ is always real.%}
  From Eq.~(\ref{FFbi6}) it follows
\be
\label{FFbi8}
a = a_0 t^{h_0}\, .
\ee
Here $a_0$ is a constant.
Then from Eqs.~(\ref{Fbi19B}), (\ref{Fbi20B}), (\ref{Fbi21B}), and
(\ref{Fbi22B}), we find
\be
\label{FFbi9}
\omega(\eta) = \frac{12\left( h_0 + n \right)^2}{\eta^2}\, ,\quad
%{\color{red}
\sigma(\zeta) = \frac{12 \left( h_0 + n \right)^2}{\zeta^2}\, ,\quad
\tilde V(\eta) = \frac{ 4 h_0^2 - 2 h_0 }{a_0^2 \eta^{2 + 2h_0} }\, ,\quad
\tilde U(\zeta) = \frac{ 4 h_0^2 - 2 h_0 }{a_0^2 \zeta^{2 + 2h_0} %}
}\, .
\ee
Thus, we have constructed the models, which describe Big Rip, quintessence, de
Sitter and  decelerating universes.
Moreover, $f$ metric seems to be dissolved by $g$
metric for such the background evolution. Note that we can present the
dynamical solution with similar properties.
Starting from $f$ universe where $b(t)=c(t)$ one can
reconstruct the evolution where these parameters just approach to $a(t)$ so
that they happen to coincide from some specific moment (for instance,
pre-inflationary stage). Such a moment will correspond to dynamical background
dissolution of $f$
metric (the perturbation theories are still different). We will not present
here the corresponding results because they are
quite complicated.

\subsubsection{Little Rip universe \label{subsubVI2b}}

The next example is Little Rip universe unifying again the description of both
of cosmologies
given by $g^\mathrm{J}_{\mu\nu}$ and $f^\mathrm{J}_{\mu\nu}$.
When $t\sim 0$, Eq.~(\ref{LR3}) indicates %{\color{red}
$\tilde H \sim - \frac{1}{t}$. %}
Then the solution of Eq.~(\ref{Fbi19B}) is given by %{\color{red}
\be
\label{LRa1}
H = - \frac{1\pm \sqrt{2}}{t}
+ \mathcal{O} \left( \frac{1}{t \left(\ln \left( \lambda g_0 t \right)\right) }
\right) \, .
\ee
Then, one gets
\be
\label{LRa3}
\omega(t) = \sigma (t) \sim \frac{12 \left(2 \pm \sqrt{2} \right)^2}{t^2}\, 
,\quad
\tilde V \sim \tilde U \sim \frac{10\pm 4\sqrt{2}}{t^2}\, .
\ee
%}
Thus, the background evolution of both of
the universes described by $g^\mathrm{J}_{\mu\nu}$ and
$f^\mathrm{J}_{\mu\nu}$ happens to coincide as Little Rip cosmology.

\subsection{General arguments for future singularity occurrence \label{subVI3}}

In Subsection \ref{subsubVI1a}, we consider the case that the finite future or
past singularity at $t=0$ in the space-time described by the metric 
$g^\mathrm{J}_{\mu\nu}$
is mapped into the infinite future or past in the space and therefore the 
singularity
is removed in the space-time described by the metric $f^\mathrm{J}_{\mu\nu}$.
This requires $\int dt \e^{\frac{\xi(t)}{2}} c(t)$ to diverge.

In this subsection, we may consider the cases that there occurs singularity in 
the
finite future or past universe described by $f^\mathrm{J}_{\mu\nu}$
but there does not occur finite future or past cosmological singularity
in the universe described by $g^\mathrm{J}_{\mu\nu}$.
\begin{itemize}
\item Even if $b(t)$ and $c(t)$ in the metric (\ref{Fbi10}) are not singular,
Eq.~(\ref{identity3}) tells $a(t)$ is not singular.
Then from (\ref{Fbi19}) and (\ref{Fbi23}), we find that $\varphi(t)$ and 
therefore
$g_{\mu\nu}$ and $g^\mathrm{J}_{\mu\nu}$ is not singular.
\item Even if $b(t)$ and $c(t)$ have a singularity, if $b(t) - c(t)$ is
non-singular, Eq.~(\ref{identity3}) shows $a(t)$ is not singular.
Then from (\ref{Fbi19}) and (\ref{Fbi23}), we find that $\varphi(t)$ and 
therefore
$g_{\mu\nu}$ and $g^\mathrm{J}_{\mu\nu}$ is not singular
although (\ref{Fbi20}) indicates that $V(t)$ is singular.
%%%%%%%%%%%%%%
%%%%%%%%%%%%%%
%%%%%%%%%%%%%%
\item Let assume $b(t) - c(t)$ has a singularity at $t=0$. Then 
(\ref{identity3}),
(\ref{Fbi19}) and (\ref{Fbi23}) demonstrate that $a(t)$ and/or $\varphi(t)$ are 
singular at
$t=0$. If $\tilde a(t)$ (\ref{FFbi1}) behaves as
\be
\label{FFFF6}
\tilde a(t) \sim \frac{l}{t}\, ,
\ee
the metric $g^\mathrm{J}_{\mu\nu}$ describes (asymptotic) de Sitter space
and therefore $g$ universe is not singular.
Using (\ref{FFbi3}), (\ref{FFFF6}), and Eq.~(\ref{Fbi19}) can be written as 
%{\color{red}
\be
\label{FFFF7}
12 \left( \frac{1}{t^2} + \frac{2H(t)}{t} + H(t)^2 \right)
= - 4 \left( \dot H - H^2 \right) - \frac{2m^2 M_\mathrm{eff}^2}{M_g^2}
\left( b(t) - c(t) \right) a(t)\, .
\ee
%}
Then one can consider the following cases:
\begin{itemize}
\item When $a(t)$ and therefore $H(t)$ are regular at $t=0$, if %{\color{red}
\be
\label{FFFF8}
b(t) - c(t) \sim - \frac{6 M_g^2}{m^2 M_\mathrm{eff}^2 a(0) } \left(
\frac{1}{t^2} + \frac{2 H(0)}{t} \right)\, ,
\ee
%}
the universe described by $g^\mathrm{J}_{\mu\nu}$ is not singular
even if the universe described by $f^\mathrm{J}_{\mu\nu}$could be singular.
\item Let assume that when $t\sim 0$, $a(t)$ behaves as $a(t) \sim a_0 t^{h_0}$
with constant $a_0$ and $h_0$. Then Eq.~(\ref{FFFF7}) demonstrates if $b(t) -
c(t)$ behaves as %{\color{red}
\be
\label{FFFF9}
b(t) - c(t) \sim \frac{2 \left( 3 + 5 h_0 + 2 h_0^2 \right) M_g^2}
{m^2 M_\mathrm{eff}^2 a_0 t^{2 - h_0}}\, ,
\ee
%}
the space-time described by $g^\mathrm{J}_{\mu\nu}$ is not singular.
\end{itemize}
\end{itemize}
It looks that all possible cases where the space-time
described by $f_{\mu\nu}$ or $f^\mathrm{J}_{\mu\nu}$ has a singularity but
there does not occur any cosmological singularity in $g$ universe are presented
in this subsection.

\section{Stability of Background Solutions}

In the previous sections, we have obtained several solutions describing the FRW 
universe.
We now discuss  the (in)stability of the obtained background solution by 
considering
the fluctuation from the background like $a\to a + \delta a$, $b\to b+ \delta 
b$,
$c \to c + \delta c$, $\varphi \to \varphi + \delta \varphi$, $\xi \to \xi + 
\delta \xi$.

By substituting Eq.~(\ref{Fbi11}) into Eq.~(\ref{Fbi12}) and using the 
constraint
(\ref{identity3}), we find
\begin{eqnarray}
\label{eq:friedmann1}
\ddot{a} - \frac{\dot{a}^2}{a} + \frac{m^2 M_\mathrm{eff}^2}{2 M_g^2} a^2
\left( b- \frac{a \dot{b}}{\dot{a}} \right)
+ \frac{1}{6} \left( 4 \dot{\varphi}^2 a - V a^3 \right) =0 \, .
\end{eqnarray}
By considering the perturbation from the background, we obtain
the linear perturbation equation from Eq.~(\ref{eq:friedmann1}):
\begin{eqnarray}
\delta \ddot{a} + A_1 \delta \dot{a} + A_2 \delta a
+ A_3 \delta \dot{b} + A_4 \delta b + A_5 \delta \dot{\varphi}
+ A_6 \delta \varphi = 0\, ,
\end{eqnarray}
where
\begin{align}
& A_1 = \frac{a^3 \dot{b} m^2 M_\mathrm{eff}^2}{2 \dot{a}^2 M_g^2}
  - \frac{2 \dot{a}}{a} \, ,\quad
A_2 =  - \frac{a^2 V}{2} + \frac{2 \dot{\varphi}^2}{3}
  - \frac{3 a^2 \dot{b} m^2 M_\mathrm{eff}^2}{2 \dot{a} M_g^2}
+ \frac{a b m^2 M_\mathrm{eff}^2}{M_g^2}
+ \frac{\dot{a}^2}{a^2} \, , \nn
& A_3 = - \frac{a^3 m^2 M_\mathrm{eff}^2}{2 \dot{a} M_g^2} \, ,\quad
A_4 = \frac{a^2 m^2 M_\mathrm{eff}^2}{2 M_g^2} \ ,\quad
A_5 = \frac{4 a \dot{\varphi}}{3} \, ,\quad
A_6 = - \frac{a^3 V'}{6} \, .
\end{align}
On the other hand, using (\ref{Fbi13}), (\ref{Fbi14}), and (\ref{identity3}), 
we obtain
\begin{eqnarray}
\dot{b}^2 \left[
  -3 \frac{M_f^2}{b^2} + m^2 M_\mathrm{eff}^2 \frac{a^2}{\dot{a}^2} \left(
1 - \frac{a^3}{b^3} \right)
+ \frac{1}{2} U \frac{a^2}{\dot{a}^2} M_f^2 \right]
+ \frac{3}{4} \dot{\xi}^2 M_f^2=0 \, .
\end{eqnarray}
Then the linear perturbation equation is given by
\begin{eqnarray}
\delta \dot{b} + B_1 \delta b + B_2 \delta \dot{a}
+ B_3 \delta a +B_4 \delta \dot{\xi} +B_5 \delta \xi = 0 \, .
\end{eqnarray}
where
\begin{align}
\label{Bs}
& B_1 = - \frac{3 M_f^2 \left ( \frac{6 M_f^2}{b^3}
+\frac{3a^5m^2M_\mathrm{eff}^2}{\dot{a}^2b^4}  \right ) \dot{\xi}^2}{ 8 \dot{b} 
\left [
\frac{a^2 M_f^2 U}{2 \dot{a}^2}- \frac{3 M_f^2}{b^2}
+ \frac{a^2 \left ( 1- \frac{a^3}{b^3}  \right )m^2 
M_\mathrm{eff}^2}{\dot{a}^2} \right ]^2} \, ,
\quad B_2 = - \frac{3 M_f^2 \dot{\xi}^2 \left ( - \frac{a^2 M_f^2 U}{\dot{a}^3}
  - \frac{2 a^2 \left ( 1-\frac{a^3}{b^3}  \right )m^2 
M_\mathrm{eff}^2}{\dot{a}^3}
\right )}{8 \dot{b} \left [
\frac{a^2 M_f^2 U}{2 \dot{a}^2} - \frac{3 M_f^2}{b^2}
+ \frac{a^2 \left (1-\frac{a^3}{b^3} \right )m^2 M_\mathrm{eff}^2}{\dot{a}^2}
\right ]^2} \, , \nn
%%%%%%%%%%%%%%%%%%%%%%%%
& B_3 = - \frac{ 3 M_f^2 \dot{\xi}^2 \left [
\frac{a M_f^2 U}{\dot{a}^2} + \frac{2 a \left( 1- \frac{a^3}{b^3}
\right )m^2 M_\mathrm{eff}^2}{\dot{a}^2} - \frac{3 a^4 m^2 
M_\mathrm{eff}^2}{\dot{a}^2 b^3} \right ]}{8 \dot{b}
\left[ \frac{a^2 M_f^2 U}{2 \dot{a}^2} - \frac{3 M_f^2}{b^2}
+ \frac{a^2 \left ( 1-\frac{a^3}{b^3}  \right )m^2 M_\mathrm{eff}^2}{\dot{a}^2}
  \right ]^2 }  \, ,  \quad
%%%%%%%%%%%%%%%%%%%%%%%%
B_4 =  - \frac{3 M_f^2 \dot{\xi}}{4 \dot{b} \left [
\frac{a^2 M_f^2 U}{2 \dot{a}^2} - \frac{3 M_f^2}{b^2}
+ \frac{a^2 \left (1-\frac{a^3}{b^3} \right )m^2 M_\mathrm{eff}^2}{\dot{a}^2} 
\right ]^2}
\, , \nn
& B_5 = - \frac{ 3 a^2 M_f^4 U' \dot{\xi}^2}{16 \dot{a}^2 \dot{b} \left [
\frac{a^2 M_f^2 U}{2 \dot{a}^2} - \frac{3 M_f^2}{b^2}
+ \frac{a^2 \left ( 1- \frac{a^3}{b^3} \right )m^2 M_\mathrm{eff}^2}{\dot{a}^2} 
\right ]^2 } \, .
\end{align}
The equation of motion for the scaler field $\varphi$ in (\ref{scalareq}) has 
the following form:
\begin{eqnarray}
\label{varphi1}
\ddot{\varphi} + 2 \frac{\dot{a}}{a} \dot{\varphi} + \frac{1}{3} V' a^2 = 0 \, 
.
\end{eqnarray}
  From this equation,  the perturbed equation for $\varphi$ is
\begin{eqnarray}
\label{varphi2}
\delta \ddot{\varphi} + C_1 \delta \dot{\varphi} + C_2 \delta \varphi
+C_3 \delta \dot{a} +C_4 \delta a =0\, ,
\end{eqnarray}
where
\be
\label{Cs}
C_1 = 2  \frac{\dot{a}}{a} \, ,\quad
C_2 = \frac{1}{3} V'' a^2 \, ,\quad
C_3 = \frac{2 \dot{\varphi}}{a} \, , \quad
C_4 = -2 \frac{\dot{a} \dot{\varphi}}{a^2} + \frac{2}{3} V' a \, .
\ee
The equation of motion for the scaler field $\xi$ in (\ref{scalareq}) has the 
following form:
\begin{eqnarray}
\label{xi1}
\ddot{\xi} + \left ( - \frac{\dot{c}}{c}+ 3 \frac{\dot{b}}{b}  \right ) 
\dot{\xi}
  + \frac{1}{3} U' c^2 =0 \, .
\end{eqnarray}
Using the constraint (\ref{identity3}), Eq.~(\ref{xi1}) can be rewritten as
\begin{eqnarray}
\label{xi2}
\ddot{\xi} + \left ( - \frac{\dot{a}}{a} + \frac{\ddot{a}}{\dot{a}}
+ 3 \frac{\dot{b}}{b} - \frac{\ddot{b}}{\dot{b}}  \right ) \dot{\xi}
+ \frac{1}{3} U' \frac{a^2 \dot{b}^2}{\dot{a}^2} = 0 \, .
\end{eqnarray}
One gets the following perturbation equation:
\begin{eqnarray}
\label{xi3}
&& \delta \ddot{\xi} + \frac{\dot{\xi}}{\dot{a}} \delta \ddot{a}
  - \frac{\dot{\xi}}{\dot{b}} \delta \ddot{b}
+ \left ( - \frac{\dot{\xi}}{a} - \frac{\ddot{a}}{\dot{a}^2} \dot{\xi}
  - \frac{2}{3} U' \frac{a^2 \dot{b}^2}{\dot{a}^3} \right ) \delta \dot{a}
+ \left (
3 \frac{\dot{\xi}}{b} + \frac{\ddot{b}}{\dot{b}^2} \dot{\xi}
+ \frac{2}{3} U'\frac{a^2 \dot{b}}{\dot{a}^2} \right ) \delta \dot{b}
+ \left ( \frac{\dot{a}}{a^2} \dot{\xi}
+ \frac{2}{3} U' \frac{a \dot{b}^2}{\dot{a}^2}
  \right ) \delta a \nonumber \\
&& - 3 \frac{\dot{b}}{b^2} \dot{\xi} \delta \dot{b}
+ \left ( - \frac{\dot{a}}{a} + \frac{\ddot{a}}{\dot{a}}
+ 3 \frac{\dot{b}}{b} - \frac{\ddot{b}}{\dot{b}} \right ) \delta \dot{\xi}
+ \frac{1}{3} U'' \frac{a^2 \dot{b}^2}{\dot{a}^2} \delta \xi = 0 \, .
\end{eqnarray}
Thus, we presented all the equations which are necessary for the study of the 
(in)stability.
The investigation of the (in)stability is, however, still tedious as it 
requests complicated numerical work.
This is because there are too many degrees
of freedom. We have $a(t)$, $b(t)$, and $c(t)$ for the metric ansatz and two 
scalar
fields $\varphi$ and $\xi$. Although we have deleted $c(t)$ by using 
(\ref{identity3}),
there are four degrees of freedom.
We need also to include their derivatives with respect to time. Then totally we 
have
eight degrees of freedom and in order to investigate the stability, we need to 
find the
eigenvalues of eight by eight matrix. Preliminary study of de Sitter space via 
these equations indicates to its stability.
We will investigate the (in)stability for several accelerating cosmological 
models in a future work.

%{\color{red}
\section{Super-luminal mode in bigravity \label{subVI4}}

In this subsection, we would like to stress that in $F(R)$ bigravity,
there may appear super-luminal mode,
that is, there can be a signal whose speed is larger than the speed of light.

The Eqs.~(\ref{Fbi10}) show that the speed $v_g$ of the massless particle which
propagates in the universe described by $g^\mathrm{J}_{\mu\nu}$ or
$g_{\mu\nu}$ is given by
$v_g^2 = \left( dx/dt \right)^2=1$ as usually in the special relativity.
Note that the light speed in the universe described by
$g^\mathrm{J}_{\mu\nu}$ is identical with the light speed in the universe
described by $g_{\mu\nu}$.
The speed $v_f$ of the massless particle which propagates in the universe
described by $f^\mathrm{J}_{\mu\nu}$ or $f_{\mu\nu}$ ,
however, is given by $v_f^2 = \left( dx/dt \right)^2=c(t)^2/b(t)^2$.
The light speed in the universe described by $f^\mathrm{J}_{\mu\nu}$
is identical with the light speed in the universe
described by $f_{\mu\nu}$, again.
Therefore if $c(t)/b(t)>1$, the speed $v_f$ is greater than the speed of light,
which propagates in $g$ universe.
When we consider the cosmology with $a(t)=b(t)=1$ in Subsection~\ref{subVI2_0},
Eq.~(\ref{FFFbi1}) shows that $c(t)>1$ except of $\tilde H=0$. Furthermore, 
because
$b(t)=1$, $v_f$ is given by
\be
\label{sl1}
v_f = 1 + \frac{ 6 \tilde H^2}{m^2 M_\mathrm{eff}^2} >1\, .
\ee
Therefore, in general, $v_f$ is greater than the speed of light
in the universe described by $g^\mathrm{J}_{\mu\nu}$ or
$g_{\mu\nu}$.
There might be no direct interaction between the matter in
the universe described by $g^\mathrm{J}_{\mu\nu}$
and the matter in the universe  described by
$f^\mathrm{J}_{\mu\nu}$
but the two kinds of matter can interact via massless and massive graviton.
This shows that if there exists any massless particle propagating
in the universe described by $f^\mathrm{J}_{\mu\nu}$ or $f_{\mu\nu}$,
the signal can propagate even in the universe described by
$g^\mathrm{J}_{\mu\nu}$ or $g_{\mu\nu}$.
The super-luminal mode can appear because $g_{\mu\nu}\neq f_{\mu\nu}$ or
$g^\mathrm{J}_{\mu\nu}\neq f^\mathrm{J}_{\mu\nu}$.
On the other hand, in the cosmology with $a(t)=b(t)=c(t)$,
because $g_{\mu\nu}= f_{\mu\nu}$ and
$g^\mathrm{J}_{\mu\nu}= f^\mathrm{J}_{\mu\nu}$, there does not appear 
super-luminal mode.
It is interesting  that the search of super-luminal particles at some era may
serve as kind of observational probe for the existence of the universe
described by $f^\mathrm{J}_{\mu\nu}$ or $f_{\mu\nu}$ (or better to say as
indication to massive gravity manifestation).

\section{Discussion}

In summary, we studied massive $F(R)$ bigravity in the conventional description
with two metrics. The variety of cosmic acceleration cosmologies is found as
explicit solution of FRW equations. In particular, Big and Little Rip, de
Sitter, quintessence and decelerating universes are constructed for Jordan
frame physical metric $g$ when Einstein frame metric $g$ is fixed and
corresponding solution for reference metric $f$ is also presented. The relation
between properties of 
$g$ and $f$ cosmologies is investigated in detail. For instance, it is
demonstrated that, in general, the physical $g$ cosmological singularity
is manifested as metric $f$ cosmological singularity.
However, there are examples where cosmological singularity of physical $g$
universe does not occur in the universe described by reference metric $f$ and
vice-versa.
Furthermore, the structure of singularity may qualitatively change: what looks
like Big Rip in one space-time manifests as Little Rip in other universe. The
solution of FRW equations where two metrics just happen to coincide is
presented for Big and Little Rip, quintessence, de Sitter and decelerating
universes. In this case, the background evolution is described via single
metric which looks quite convenient even keeping in mind that second metric is
just the effective description of exotic matter (the perturbation theory for
two metrics is also different). Perturbation theory for cosmological solutions 
under discussions is also developed.

We also observed that the massless particle in the space-time given by the
metric $f_{\mu\nu}$ or $f^\mathrm{J}_{\mu\nu}$ can be super-luminal.
Then if there appears any indication that there exists a signal whose speed is
greater than the speed of light, it may be the indication for another
space-time existence (massive graviton effect).

It could be interesting to consider how we can observe the manifestation of the
space-time described by the metric $f_{\mu\nu}$ or $f^\mathrm{J}_{\mu\nu}$.
As we mentioned, we can observe the matter via massless and massive
gravitons. Therefore not only dark energy but also dark matter might be a
matter in the space-time given by the metric $f_{\mu\nu}$ or
$f^\mathrm{J}_{\mu\nu}$.

\section*{Acknowledgments.}

We are grateful to G. Gabadadze, N. Kaloper, M. Sami, A. Starobinsky, J.Soda
and K. Bamba for useful discussions.
The work by SN is supported in part by Global COE Program of Nagoya University
(G07) provided by the Ministry of Education, Culture, Sports, Science \&
Technology and by the JSPS Grant-in-Aid for Scientific Research (S) \# 22224003
and (C) \# 23540296.
The work by SDO is supported in part by MICINN (Spain),  project
FIS2010-15640 and by AGAUR (Generalitat de Ca\-ta\-lu\-nya),
contract 2009SGR-994.


\begin{thebibliography}{99}

%\cite{Fierz:1939ix}
\bibitem{Fierz:1939ix}
M.~Fierz and W.~Pauli,
%``On relativistic wave equations for particles of
%arbitrary spin in an electromagnetic field,''
Proc.\ Roy.\ Soc.\ Lond.\ A {\bf 173} (1939) 211.
%%CITATION = PRSLA,A173,211;%%

%\cite{Hinterbichler:2011tt}
\bibitem{Hinterbichler:2011tt}
K.~Hinterbichler,
%``Theoretical Aspects of Massive Gravity,''
Rev.\ Mod.\ Phys.\  {\bf 84} (2012) 671
[arXiv:1105.3735 [hep-th]].
%%CITATION = ARXIV:1105.3735;%%

%\cite{Boulware:1974sr}
\bibitem{Boulware:1974sr}
D.~G.~Boulware and S.~Deser,
%``Classical General Relativity Derived from Quantum Gravity,''
Annals Phys.\  {\bf 89} (1975) 193.
%%CITATION = APNYA,89,193;%%

%\cite{vanDam:1970vg}
\bibitem{vanDam:1970vg}
H.~van Dam and M.~J.~G.~Veltman,
%``Massive and massless Yang-Mills and gravitational fields,''
Nucl.\ Phys.\ B {\bf 22} (1970) 397; \\
%%CITATION = NUPHA,B22,397;%%
%\cite{Zakharov:1970cc}
%\bibitem{Zakharov:1970cc}
V.~I.~Zakharov,
%``Linearized gravitation theory and the graviton mass,''
JETP Lett.\  {\bf 12} (1970) 312
[Pisma Zh.\ Eksp.\ Teor.\ Fiz.\  {\bf 12} (1970) 447].
%%CITATION = JTPLA,12,312;%%

%\cite{Vainshtein:1972sx}
\bibitem{Vainshtein:1972sx}
A.~I.~Vainshtein,
%``To the problem of nonvanishing gravitation mass,''
Phys.\ Lett.\ B {\bf 39} (1972) 393.
%%CITATION = PHLTA,B39,393;%%

%\cite{Luty:2003vm}
\bibitem{Luty:2003vm}
M.~A.~Luty, M.~Porrati and R.~Rattazzi,
%``Strong interactions and stability in the DGP model,''
JHEP {\bf 0309}, 029 (2003)
[hep-th/0303116]; \\
%%CITATION = HEP-TH/0303116;%%
%\cite{Nicolis:2004qq}
%\bibitem{Nicolis:2004qq}
A.~Nicolis and R.~Rattazzi,
%``Classical and quantum consistency of the DGP model,''
JHEP {\bf 0406} (2004) 059  [hep-th/0404159].
%%CITATION = HEP-TH/0404159;%%

%\cite{Dvali:2000hr}
\bibitem{Dvali:2000hr}
G.~R.~Dvali, G.~Gabadadze and M.~Porrati,
%``4-D gravity on a brane in 5-D Minkowski space,''
Phys.\ Lett.\ B {\bf 485} (2000) 208
[hep-th/0005016]; \\
%%CITATION = HEP-TH/0005016;%%
%\cite{Deffayet:2000uy}
%\bibitem{Deffayet:2000uy}
C.~Deffayet,
%``Cosmology on a brane in Minkowski bulk,''
Phys.\ Lett.\  B {\bf 502}, 199 (2001)
[arXiv:hep-th/0010186]; \\
%%CITATION = PHLTA,B502,199;%%
%\cite{Deffayet:2001pu}
%\bibitem{Deffayet:2001pu}
C.~Deffayet, G.~R.~Dvali and G.~Gabadadze,
%``Accelerated universe from gravity leaking to extra dimensions,''
Phys.\ Rev.\ D {\bf 65}, 044023 (2002)
[astro-ph/0105068].
%%CITATION = ASTRO-PH/0105068;%%

%\cite{deRham:2010ik}
\bibitem{deRham:2010ik}
C.~de Rham and G.~Gabadadze,
%``Generalization of the Fierz-Pauli Action,''
Phys.\ Rev.\ D {\bf 82}, 044020 (2010)
[arXiv:1007.0443 [hep-th]]; \\
%%CITATION = ARXIV:1007.0443;%%
%\cite{deRham:2010kj}
%\bibitem{deRham:2010kj}
C.~de Rham, G.~Gabadadze and A.~J.~Tolley,
%``Resummation of Massive Gravity,''
Phys.\ Rev.\ Lett.\  {\bf 106} (2011) 231101
[arXiv:1011.1232 [hep-th]].
%%CITATION = ARXIV:1011.1232;%%

%\cite{Hassan:2011hr}
\bibitem{Hassan:2011hr}
S.~F.~Hassan and R.~A.~Rosen,
%``Resolving the Ghost Problem in non-Linear Massive Gravity,''
Phys.\ Rev.\ Lett.\  {\bf 108} (2012) 041101
[arXiv:1106.3344 [hep-th]].
%%CITATION = ARXIV:1106.3344;%%

%\cite{Hassan:2011zd}
\bibitem{Hassan:2011zd}
S.~F.~Hassan and R.~A.~Rosen,
%``Bimetric Gravity from Ghost-free Massive Gravity,''
JHEP {\bf 1202} (2012) 126
[arXiv:1109.3515 [hep-th]].
%%CITATION = ARXIV:1109.3515;%%

%%%%%%%%%%%%%%%%%%%%%%%%%%

%\cite{Hassan:2011tf}
\bibitem{Hassan:2011tf}
S.~F.~Hassan, R.~A.~Rosen and A.~Schmidt-May,
%``Ghost-free Massive Gravity with a General Reference Metric,''
JHEP {\bf 1202} (2012) 026
[arXiv:1109.3230 [hep-th]].
%%CITATION = ARXIV:1109.3230;%%
%77 citations counted in INSPIRE as of 20 Apr 2013

%\cite{Hassan:2011vm}
\bibitem{Hassan:2011vm}
S.~F.~Hassan and R.~A.~Rosen,
%``On Non-Linear Actions for Massive Gravity,''
JHEP {\bf 1107} (2011) 009
[arXiv:1103.6055 [hep-th]].
%%CITATION = ARXIV:1103.6055;%%
%76 citations counted in INSPIRE as of 20 Apr 2013

%%%%%%%%%%%%%%%%%%%%%%%%%%

%\cite{Golovnev:2011aa}
\bibitem{Golovnev:2011aa}
A.~Golovnev,
%``On the Hamiltonian analysis of non-linear massive gravity,''
Phys.\ Lett.\ B {\bf 707} (2012) 404
[arXiv:1112.2134 [gr-qc]]; \\
%%CITATION = ARXIV:1112.2134;%%
%\cite{Kluson:2012wf}
%\bibitem{Kluson:2012wf}
J.~Kluson,
%``Non-Linear Massive Gravity with Additional Primary
%Constraint and Absence of Ghosts,''
arXiv:1204.2957 [hep-th];
arXiv:1209.3619;
%%CITATION = ARXIV:1204.2957;%%
%\cite{Hassan:2012qv}
%\bibitem{Hassan:2012qv}
S.~F.~Hassan, A.~Schmidt-May and M.~von Strauss,
%``Proof of Consistency of Nonlinear Massive Gravity
%in the St\'uckelberg Formulation,''
arXiv:1203.5283 [hep-th]; \\
%%CITATION = ARXIV:1203.5283;%%
%\cite{Hassan:2011ea}
%\bibitem{Hassan:2011ea}
S.~F.~Hassan and R.~A.~Rosen,
%``Confirmation of the Secondary Constraint and Absence of
%Ghost in Massive Gravity and Bimetric Gravity,''
JHEP {\bf 1204} (2012) 123
[arXiv:1111.2070 [hep-th]]; \\
%%CITATION = ARXIV:1111.2070;%%
%%
%\cite{Koyama:2011yg}
%\bibitem{Koyama:2011yg}
K.~Koyama, G.~Niz and G.~Tasinato,
%``Strong interactions and exact solutions in non-linear massive gravity,''
Phys.\ Rev.\ D {\bf 84} (2011) 064033
[arXiv:1104.2143 [hep-th]]; arXiv:1210.4378;\\
%%CITATION = ARXIV:1104.2143;%%
%%
%\cite{D'Amico:2011jj}
%\bibitem{D'Amico:2011jj}
G.~D'Amico, C.~de Rham, S.~Dubovsky, G.~Gabadadze, D.~Pirtskhalava and
A.~J.~Tolley,
%``Massive Cosmologies,''
Phys.\ Rev.\ D {\bf 84} (2011) 124046
[arXiv:1108.5231 [hep-th]]; \\
%%CITATION = ARXIV:1108.5231;%%
%%
%\cite{Hinterbichler:2012cn}
%\bibitem{Hinterbichler:2012cn}
K.~Hinterbichler and R.~A.~Rosen,
%``Interacting Spin-2 Fields,''
JHEP {\bf 1207} (2012) 047
[arXiv:1203.5783 [hep-th]]; \\
%%CITATION = ARXIV:1203.5783;%%
%%
%\cite{Baccetti:2012bk}
%\bibitem{Baccetti:2012bk}
V.~Baccetti, P.~Martin-Moruno and M.~Visser,
%``Massive gravity from bimetric gravity,''
arXiv:1205.2158 [gr-qc]; \\
%%CITATION = ARXIV:1205.2158;%%
%%
%\cite{Kobayashi:2012fz}
%\bibitem{Kobayashi:2012fz}
T.~Kobayashi, M.~Siino, M.~Yamaguchi and D.~Yoshida,
%``New Cosmological Solutions in Massive Gravity,''
arXiv:1205.4938 [hep-th]; \\
%%CITATION = ARXIV:1205.4938;%%
%\cite{Nomura:2012xr}
%\bibitem{Nomura:2012xr}
K.~Nomura and J.~Soda,
%``When is Multimetric Gravity Ghost-free?,''
Phys.\ Rev.\ D {\bf 86} (2012) 084052
[arXiv:1207.3637 [hep-th]]; \\
%%CITATION = ARXIV:1207.3637;%%
%\cite{Saridakis:2012jy}
%\bibitem{Saridakis:2012jy}
E.~N.~Saridakis,
%``Phantom crossing and quintessence limit in extended nonlinear massive
%gravity,''
arXiv:1207.1800 [gr-qc]; \\
%%CITATION = ARXIV:1207.1800;%%
%\cite{Cai:2012ag}
%\bibitem{Cai:2012ag}
Y.~-F.~Cai, C.~Gao and E.~N.~Saridakis,
%``Bounce and cyclic cosmology in extended nonlinear massive gravity,''
JCAP {\bf 1210} (2012) 048
[arXiv:1207.3786 [astro-ph.CO]]; \\
Y. Zhang, R. Saito and M. Sasaki, arXiv:1210.6224.
%%CITATION = ARXIV:1207.3786;%%
%\cite{Mohseni:2012ug}
%\bibitem{Mohseni:2012ug}
M.~Mohseni,
%``Gravitational Waves in Ghost Free Bimetric Gravity,''
JCAP {\bf 1211} (2012) 023
[arXiv:1211.3501 [hep-th]].
%%CITATION = ARXIV:1211.3501;%%

%\cite{Damour:2002wu}
\bibitem{Damour:2002wu}
T.~Damour, I.~I.~Kogan and A.~Papazoglou,
%``Nonlinear bigravity and cosmic acceleration,''
Phys.\ Rev.\ D {\bf 66} (2002) 104025
[hep-th/0206044].
%%CITATION = HEP-TH/0206044;%%

%\cite{Volkov:2011an}
\bibitem{Volkov:2011an}
M.~S.~Volkov,
%``Cosmological solutions with massive gravitons in the bigravity theory,''
JHEP {\bf 1201} (2012) 035
[arXiv:1110.6153 [hep-th]];arXiv:1207.3723.
%%CITATION = ARXIV:1110.6153;%%

%%%%%%%%%%%%%%
%%%%%%%%%%%%%%

%\cite{vonStrauss:2011mq}
\bibitem{vonStrauss:2011mq}
M.~von Strauss, A.~Schmidt-May, J.~Enander, E.~Mortsell, S.~F.~Hassan and ,
%``Cosmological Solutions in Bimetric Gravity and their Observational Tests,''
JCAP {\bf 1203} (2012) 042
[arXiv:1111.1655 [gr-qc]].
%%CITATION = ARXIV:1111.1655;%%
%45 citations counted in INSPIRE as of 30 Mar 2013

%\cite{Volkov:2012cf}
\bibitem{Volkov:2012cf}
M.~S.~Volkov,
%``Exact self-accelerating cosmologies in the ghost-free bigravity and massive 
%gravity,''
Phys.\ Rev.\ D {\bf 86} (2012) 061502
[arXiv:1205.5713 [hep-th]].
%%CITATION = ARXIV:1205.5713;%%
%21 citations counted in INSPIRE as of 30 Mar 2013

%\cite{Berg:2012kn}
\bibitem{Berg:2012kn}
M.~Berg, I.~Buchberger, J.~Enander, E.~Mortsell, S.~Sjors and ,
%``Growth Histories in Bimetric Massive Gravity,''
JCAP {\bf 1212} (2012) 021
[arXiv:1206.3496 [gr-qc]].
%%CITATION = ARXIV:1206.3496;%%
%16 citations counted in INSPIRE as of 30 Mar 2013

%%%%%%%%%%%%%%
%%%%%%%%%%%%%%

%\cite{Nojiri:2012zu}
\bibitem{Nojiri:2012zu}
S.~Nojiri and S.~D.~Odintsov,
%``Ghost-free $F(R)$ bigravity and accelerating cosmology,''
arXiv:1207.5106 [hep-th].
%%CITATION = ARXIV:1207.5106;%%

%\cite{Akrami:2012vf}
\bibitem{Akrami:2012vf}
Y.~Akrami, T.~S.~Koivisto and M.~Sandstad,
%``Accelerated expansion from ghost-free bigravity:
%a statistical analysis with improved generality,''
arXiv:1209.0457 [astro-ph.CO].
%%CITATION = ARXIV:1209.0457;%%


%\cite{Nojiri:2006ri}
\bibitem{Nojiri:2006ri}
S.~Nojiri and S.~D.~Odintsov,
%``Introduction to modified gravity and gravitational alternative for dark
% energy,''
eConf C {\bf 0602061} (2006) 06
[Int.\ J.\ Geom.\ Meth.\ Mod.\ Phys.\  {\bf 4} (2007) 115]
[hep-th/0601213]; \\
%%CITATION = HEP-TH/0601213;%%
%%%%%%%%%
%\cite{Nojiri:2010wj}
%\bibitem{Nojiri:2010wj}
S.~Nojiri and S.~D.~Odintsov,
%``Unified cosmic history in modified gravity: from F(R)
%theory to Lorentz non-invariant models,''
Phys.\ Rept.\  {\bf 505} (2011) 59
[arXiv:1011.0544 [gr-qc]].
%%CITATION = ARXIV:1011.0544;%%

%\cite{Capozziello:2010zz}
\bibitem{Capozziello:2010zz}
S.~Capozziello and V.~Faraoni,
``Beyond Einstein gravity: A Survey of gravitational theories
for cosmology and astrophysics,'',
DOI: 10.1007/978-94-007-0165-6; \\
%%CITATION = INSPIRE-1107700;%%
%\cite{Capozziello:2011et}
%\bibitem{Capozziello:2011et}
S.~Capozziello and M.~De Laurentis,
%``Extended Theories of Gravity,''
Phys.\ Rept.\  {\bf 509} (2011) 167
[arXiv:1108.6266 [gr-qc]].
%%CITATION = ARXIV:1108.6266;%%

%\cite{Nojiri:2003ft}
\bibitem{Nojiri:2003ft}
S.~Nojiri and S.~D.~Odintsov,
%``Modified gravity with negative and positive powers of the curvature:
%Unification of the inflation and of the cosmic acceleration,''
Phys.\ Rev.\ D {\bf 68} (2003) 123512
[hep-th/0307288].
%%CITATION = HEP-TH/0307288;%%

%\cite{Caldwell:1999ew}
\bibitem{Caldwell:1999ew}
R.~R.~Caldwell,
%``A Phantom menace?,''
Phys.\ Lett.\ B {\bf 545} (2002) 23
[astro-ph/9908168]; \\
%%CITATION = ASTRO-PH/9908168;%%
%\cite{Starobinsky:1999yw}
%\bibitem{Starobinsky:1999yw}
A.~A.~Starobinsky,
%``Future and origin of our universe: Modern view,''
Grav.\ Cosmol.\  {\bf 6} (2000) 157
[astro-ph/9912054].
%%CITATION = ASTRO-PH/9912054;%%


%\cite{Caldwell:2003vq}
\bibitem{Caldwell:2003vq}
R.~R.~Caldwell, M.~Kamionkowski and N.~N.~Weinberg,
%``Phantom energy and cosmic doomsday,''
Phys.\ Rev.\ Lett.\  {\bf 91} (2003) 071301
[astro-ph/0302506].
%%CITATION = ASTRO-PH/0302506;%%

%\cite{Frampton:2011sp}
\bibitem{Frampton:2011sp}
P.~H.~Frampton, K.~J.~Ludwick and R.~J.~Scherrer,
%``The Little Rip,''
Phys.\ Rev.\ D {\bf 84} (2011) 063003
[arXiv:1106.4996 [astro-ph.CO]].
%%CITATION = ARXIV:1106.4996;%%

%\cite{Brevik:2011mm}
\bibitem{Brevik:2011mm}
I.~Brevik, E.~Elizalde, S.~Nojiri and S.~D.~Odintsov,
%``Viscous Little Rip Cosmology,''
Phys.\ Rev.\ D {\bf 84} (2011) 103508
[arXiv:1107.4642 [hep-th]].
%%CITATION = ARXIV:1107.4642;%%

%\cite{Frampton:2011rh}
\bibitem{Frampton:2011rh}
P.~H.~Frampton, K.~J.~Ludwick, S.~Nojiri, S.~D.~Odintsov and R.~J.~Scherrer,
%``Models for Little Rip Dark Energy,''
Phys.\ Lett.\ B {\bf 708} (2012) 204
[arXiv:1108.0067 [hep-th]].
%%CITATION = ARXIV:1108.0067;%%

%\cite{Capozziello:2012re}
\bibitem{Capozziello:2012re}
S.~Capozziello and P.~Martin-Moruno,
%``Bounces, turnarounds and singularities in bimetric gravity,''
arXiv:1211.0214 [gr-qc].

%\cite{Amanullah:2010vv}
\bibitem{Amanullah:2010vv}
R.~Amanullah, C.~Lidman, D.~Rubin, G.~Aldering, P.~Astier,
K.~Barbary, M.~S.~Burns and A.~Conley {\it et al.},
%``Spectra and Light Curves of Six Type Ia Supernovae at 0.511 < z < 1.12
%and the Union2 Compilation,''
Astrophys.\ J.\  {\bf 716} (2010) 712
[arXiv:1004.1711 [astro-ph.CO]].
%%CITATION = ARXIV:1004.1711;%%

%\cite{BezerradeMello:2012nq}
\bibitem{BezerradeMello:2012nq}
E.~R.~Bezerra de Mello, A.~A.~Saharian and ,
%``Scalar self-energy for a charged particle in global monopole spacetime with 
a spherical boundary,''
Class.\ Quant.\ Grav.\  {\bf 29} (2012) 135007
[arXiv:1201.1770 [hep-th]].
%%CITATION = ARXIV:1201.1770;%%
%1 citations counted in INSPIRE as of 31 Mar 2013

%\cite{Deser:2012qx}
%\bibitem{Deser:2012qx}
%S.~Deser, A.~Waldron and ,
%``Acausality of Massive Gravity,''
%Phys.\  Rev.\  Lett.\  110, {\bf 111101} (2013)
%[arXiv:1212.5835 [hep-th]].
%%CITATION = ARXIV:1212.5835;%%
%13 citations counted in INSPIRE as of 31 Mar 2013

%\cite{deRham:2011pt}
\bibitem{deRham:2011pt}
C.~de Rham, G.~Gabadadze, A.~J.~Tolley and ,
%``Comments on (super)luminality,''
arXiv:1107.0710 [hep-th].
%%CITATION = ARXIV:1107.0710;%%
%20 citations counted in INSPIRE as of 31 Mar 2013





\end{thebibliography}
\end{document}